\documentclass[twocolumn,showpacs,preprintnumbers,amsmath,amssymb]{revtex4}

\usepackage{amssymb}
\usepackage{mathrsfs}

\usepackage{graphicx}
\usepackage{dcolumn}
\usepackage{bm}
\usepackage{mathrsfs}

\begin{document}

\title{Assessing the discovery 
potential of directional detection of Dark Matter}

\author{J. Billard}
\email{billard@lpsc.in2p3.fr}
  
\author{F. Mayet}

\author{D. Santos}
\affiliation{Laboratoire de Physique Subatomique et de Cosmologie, Universit\'e Joseph Fourier Grenoble 1,
  CNRS/IN2P3, Institut Polytechnique de Grenoble, Grenoble, France}

%
%
\date{\today}

\begin{abstract}
There is a worldwide effort toward the development of a large TPC (Time Projection Chamber) devoted to directional Dark Matter detection.  All current projects are being 
designed to fulfill a unique goal : identifying weakly interacting massive particle (WIMP)  as such by taking advantage of the expected 
direction dependence of WIMP-induced events toward the constellation Cygnus. However such proof of discovery requires a careful statistical data treatment. In this paper, the discovery potential of forthcoming 
directional detectors is adressed by  using a
frequentist approach based on the profile likelihood ratio test statistic. This allows us to estimate the expected significance of a Dark Matter detection taking into account
astrophysical and experimental uncertainties.
We show that the energy threshold and the background contamination are key experimental issues for directional detection, while angular resolution and sense recognition
efficiency only
mildly affect the sensitivity and the energy resolution is unimportant.
This way, we found that a 30 kg.year CF$_4$ directional experiment could reach 
a 3$\sigma$ sensitivity at 90\% C.L. down to  
$10^{-5}$ pb and $3.10^{-4}$ pb for the WIMP-proton axial cross section in the most optimistic and pessimistic detector performance case respectively. 
\end{abstract}

\maketitle

%

\section{Introduction}
An ever increasing body of evidence supports the existence of Cold Dark Matter (CDM) as the major contribution to 
the matter budget of the Universe. On the largest scale, cosmological measurements \cite{wmap} tightly constrain the CDM relic 
density whereas on a local scale, the discrepancy between the rotation curves of spiral galaxies, as measured by 21 cm observations, and the ones
 that would be inferred from the luminous matter, indicates that 
spiral galaxies, including the Milky Way,  should be embedded in a Dark Matter halo~\cite{halo,halo.mw}.\\
Directional detection of galactic Dark Matter has been   proposed  as  a powerful tool
to identify genuine WIMP events as such \cite{spergel}. 
Indeed,  one expects a clear asymmetry in the angular distribution of WIMP-induced events in the direction of motion of the 
Solar system, which  happens to be roughly in the direction of the  constellation Cygnus. As the background distribution is expected to be isotropic in the galactic restframe, one expects a 
 clear and unambiguous difference between the WIMP-induced signal and the background one. Recent studies have shown that, within the framework of dedicated statistical data analysis, a
low exposure directional detector could lead either to a high significance discovery
of galactic Dark Matter \cite{billard.disco,green.disco,billard.identification} or to a competitive exclusion  \cite{billard.exclusion}.\\
Non-directional direct detection of Dark Matter is entering in a new era as several detectors are starting to exclude the upper part 
of the predicted supersymmetric parameter space, either in the spin-independent channel \cite{Ahmed:2011gh,Aprile:2011hi,Aalseth:2010vx,Angloher:2011uu} or in the spin-dependent one 
\cite{simple,edelweiss-spin,coupp,kims,naiad,picasso,xenon10,zeplin}.  On the other hand, forthcoming directional detection projects 
are designed to claim a proof of discovery. Both approaches require a dedicated statistical treatment, taking into account systematics related to detector performance 
as well as astrophysical uncertainties.  Following  \cite{Cowan:2010js,Aprile:likelihood}, we use a profile likelihood ratio test statistic to assess the discovery potential of directional detection.\\
We give a complete overview of the effect of the main experimental issues on the discovery potential of directional detection of Dark Matter. Our
aim is to establish a weighted wish list that could be used when trying to optimize the design of a Dark Matter directional detection experiment. 
The experimental issues considered hereafter are: the background contamination, the energy threshold, the sense recognition efficiency, the angular
resolution, the energy resolution and the background modeling. With a profile likelihood method, we estimate the sensitivity of a given directional
experiment by taking into account the most relevant astrophysical uncertainties, {\it e.g.} the local Dark Matter density, the WIMP velocity distribution and the 
velocity of the Solar System orbit.\\
The paper is organised as follows. Section~\ref{sec:framework} presents a brief
introduction of directional Dark Matter detection, both from an experimental point
of view and from a theoretical one. Section~\ref{sec:profile} introduces the definition of the likelihood function we have used, 
as well as the formalism of a frequentist approach to estimate the significance with a profile likelihood ratio test statistic. Section~\ref{sec:effect} focuses on the
application of the profile likelihood test to estimate the impact of the different experimental issues on the sensitivity of upcoming directional detector.
 Finally, in section~\ref{sec:prospect} we conclude this study by considering two very different detectors with their respective performance to evaluate both their
sensitivity to Dark Matter from supersymmetric models and their competitivity in comparison to  existing limits.


\section{Directional detection framework}
\label{sec:framework} 
\subsection{Detector configuration}
\label{sec:dec}
There is a worldwide effort toward the development of a large TPC (Time Projection Chamber) devoted to directional detection \cite{white} and 
all current projects \cite{dmtpc,drift,d3,emulsions,mimac,newage} 
face common experimental challenges and share a unique goal :  the simultaneous measurement of the energy ($E_r$) and the 3D track ($\Omega_r$) of low energy recoils, 
thus  allowing to evaluate the double-differential spectrum $\mathrm{d}^2R/\mathrm{d}E_r\mathrm{d}\Omega_r$ down to the energy threshold. 
This can be achieved with low pressure gaseous detectors (TPC) and several gases have been suggested : 
$\rm CF_4$, $\rm ^{3}He$, $\rm C_4H_{10}$ or $\rm CS_2$. As a matter of fact, experimental key issues include : sense recognition~\cite{dujmic,burgos,majewski}, 
angular and energy resolutions, energy threshold  as well as the 
residual background contamination.\\ 
Bearing in mind the need for a directional detector optimization~\cite{billard.exclusion,Green:2006cb} at the current  
stage of detector design, we present in section \ref{sec:effect} a study of the effect of detector configurations in the case of a working example : 
a low exposure (30 kg.year)  $\rm CF_4$ TPC, operated at low pressure and  allowing  3D track reconstruction as proposed by the MIMAC collaboration \cite{mimac}.

\subsection{Directional detection}
\label{sec:directional}
Directional detection depends crucially on the local WIMP velocity distribution \cite{alenazi.directionnelle,Green:2010gw,Serpico:2010ae}. The isothermal sphere halo model is 
often considered but it is worth going beyond this standard paradigm when trying to account for all astrophysical uncertainties. 
The multivariate Gaussian WIMP velocity distribution corresponds to the generalization of the standard isothermal sphere \cite{Evans:2000gr} with a density profile 
$\rho(r)\propto 1/r^2$, leading to a smooth WIMP velocity distribution,  
a flat rotation curve and no substructure. The WIMP velocity distribution in the laboratory frame is   given by,
\begin{equation}
f(\vec{v}) = \frac{1}{(8\pi^3\det{\boldsymbol\sigma}^2_v)^{1/2}}\exp{\left[-\frac{1}{2}(\vec{v} - \vec{v}_{\odot})^T {\boldsymbol\sigma}^{-2}_v(\vec{v} - \vec{v}_{\odot})\right]}
\end{equation}
where ${\boldsymbol\sigma}_v = \text{diag}[\sigma_{x}, \sigma_{y}, \sigma_{z}]$ is the velocity dispersion tensor 
assumed to be diagonal in the Galactic rest frame ($\hat{x}$, $\hat{y}$, $\hat{z}$) and $\vec{v}_{\odot}$ is the Sun's velocity vector with respect to
the Galactic rest frame. When neglecting the Sun peculiar velocity and the Earth orbital 
velocity about the Sun,  $\vec{v}_{\odot}$ corresponds to the detector velocity in
the Galactic rest frame and is taken to be $v_{\odot} = 220$ km.s$^{-1}$ along the $\hat{y}$ axis pointing toward the Cygnus constellation at 
($\ell_{\odot} = 90^{\circ}$, $b_{\odot} = 0^{\circ}$).\\
The  directional recoil  rate  is given by  \cite{gondolo} :
\begin{equation}
\frac{\mathrm{d}^2R}{\mathrm{d}E_r\mathrm{d}\Omega_r} = \frac{\rho_0\sigma_0}{4\pi m_{\chi}m^2_r}F^2(E_r)\hat{f}(v_{\text{min}},\hat{q}),
\label{directionalrate}
\end{equation}
with $m_{\chi}$ the WIMP mass, $m_r$ the WIMP-nucleus reduced mass, $\rho_0$   the local Dark Matter density, $\sigma_0$   the
WIMP-nucleus elastic scattering cross section, $F(E_r)$  the form factor  (using the axial expression from \cite{lewin}),  
$v_{\text{min}}$ the   minimal WIMP velocity required to produce a
nuclear recoil of energy $E_r$ and $\hat{q}$ the direction of the recoil momentum. 
Finally, $\hat{f}(v_{\text{min}},\hat{q})$ is the three-dimensional Radon transform of the WIMP 
velocity distribution $f(\vec{v})$.  Using the Fourier slice theorem \cite{gondolo},    the Radon transform of the 
multivariate 
Gaussian is,
\begin{equation}
\hat{f}(v_{\text{min}},\hat{q}) = \frac{1}{(2\pi\hat{q}^T{\boldsymbol\sigma}^2_v\hat{q})^{1/2}}\exp{\left[-\frac{\left[v_{\text{min}} - \hat{q}.\vec{v}_{\odot}\right]^2}{2\hat{q}^T{\boldsymbol\sigma}^2_v\hat{q}}\right]}.
\end{equation}

As outlined in \cite{billard.disco,billard.identification,green.disco}, a clear and unambiguous signature in favor of a Dark Matter detection is given by the fact that the   
 recovered main recoil 
direction is shown to be pointing toward Cygnus within a few degrees. However, to assess the significance of the discovery, in a frequentist approach and taking into
account astrophysical uncertainties, a complete statistical analysis is required, as outlined below.

\subsection{Astrophysical uncertainties}
\label{sec:astro.def}
The effect of astrophysical parameters on exclusion limits  deduced from  direct detection, both from energy, annual modulation and direction measurements, has been investigated in detail 
in~\cite{billard.exclusion,nezri,Green:2010gw,McCabe:2010zh,Kuhlen:2009vh,Fox:2010bz,Fox:2010bu,Strigari:2009zb,Serpico:2010ae}. 
For a complete discussion on their evaluations, uncertainties  and  effect on dark matter detection, we refer the reader to 
\cite{Cerdeno:2010jj,Green:2010ri} and references therein.
For dark matter search, both direct and directional, it is indeed of great 
interest to account for   uncertainties on astrophysical parameters, as in \cite{Aprile:likelihood}, although it has been done for the
escape velocity only, within the framework of a   profile likelihood ratio test statistic.\\
Three astrophysical parameters play a key role : the local dark matter density $\rho_0$, the Sun's velocity vector $v_\odot$ and the WIMP velocity 
distribution, parametrized by the velocity dispersions along the three axes $\sigma_{x,y,z}$.\\

As the WIMP event rate is proportional to  the quantity $\rho_0 \times \sigma_p$, the  
value of the local dark matter density $\rho_0$ directly affects the estimation of the WIMP-proton cross section $\sigma_p$. 
A standard value $\rm \rho_0=0.3 \   {\rm GeV.c^{-2}}$ is usually used  for the sake of comparison of  various direct detector results  \cite{Gates,pdg}.
However, there are still debates on the value of $\rho_0$ and several papers, using different techniques, have  found the following  values of  the local Dark Matter density : 
$\rho_0 ({\rm  GeV/c^2/cm^3})$ equal to
$0.32\pm 0.07$ \cite{Strigari:2009zb},  
$0.43  \pm 0.11 \pm 0.10$ \cite{salucci}, 
$0.3   \pm 0.1$ \cite{Weber:2009pt}, 
and $0.39 \pm 0.03$ \cite{Catena:2009mf}. Clearly, systematic errors arise from uncertainties in modeling the Milky Way. 
Interestingly,  
S. Garbari {\it et al.} \cite{Garbari:2011dh} point out that it is vital to measure the vertical dispersion profile of the tracers to recover an
unbiased estimate of $\rm \rho_0$. Depending on the choice of the velocity distribution of star tracer population (from isothermal to 
non-isothermal), their estimation of $\rm \rho_0$ can vary by a factor of 10 or so. Recently, 
M.~C.~Smith {\it et al.} \cite{Smith:2011fs} have  computed a local dark matter
density of $0.57 \ {\rm  GeV/c^2/cm^3}$ using SDSS data \cite{Bramich} by studying the kinematics of Galactic disk stars in the solar
neighborhood.\\
Valuable information are also extracted from recent high resolution N-body simulations of Milky Way like objects.
In particular, it is shown in \cite{Pato:2010yq} that the estimation of the local Dark Matter density, 
on the stellar disk and at $\sim$8 kpc from the Galactic center, is found to be always larger, by 20\% or so,  
than the average density in a spherical shell of same radius, {\it i.e} the above quoted quantity 
inferred from dynamical measurements.  Interestingly, the local Dark Matter density 
distribution   is predicted \cite{vogel} to be remarkably smooth, varying by less than 15\% at
the 99.9\% C.L.  from the average value over an ellipsoidal shell. This highlights the fact that the local density, at the Sun's location, 
should not differ from the above estimations.\\
These arguments, both from observations and simulations,  add still more weights to the need of taking a possible range of values for the local Dark Matter density rather than a fixed  
standard value ($\rm \rho_0=0.3 \   {\rm GeV.c^{-2}}$), although it allows a fair comparison of various dark matter search results.
Hence, to account for all estimations and uncertainties, the local Dark Matter density  $\rm \rho_0$ is treated hereafter as a nuisance parameter 
in the following, considering the range  $0.3\pm 0.1 \ {\rm  GeV/c^2/cm^3}$, see tab.~\ref{tab:prior}.\\
On the choice of mean value and uncertainty, it is worth emphasizing that the local Dark Matter density is indeed a key 
experimental issue for directional Dark Matter search (as for direction-insensitive one) but its effect is 
rather simple to handle. For instance, taking a larger value ({\it e.g.} as proposed by \cite{Smith:2011fs}), say 0.4 instead of 0.3, will mainly change the sensitivity by 33\%, giving in fact a 
sensitivity to smaller cross sections. For a study aiming at showing the reach of directional detection, a modification of the range chosen
for the local Dark Matter density does not affect the result. This is not the case however, when presenting an experimental result as
oulined above.\\

The second astrophysical parameter to be carefully handled is the Sun's velocity $v_\odot$ equal to the local circular speed when neglecting the Sun's peculiar velocity.
The standard value is $220 \pm 20 \ {\rm km.s^{-1}}$ \cite{kerr}. As outlined in \cite{Cerdeno:2010jj,McCabe:2010zh,Green:2010ri}, recent determinations of 
its value span on a wide range and the impact on directionality has been studied~\cite{billard.exclusion,copi2}. 
Several papers, using different techniques, have  studied the value of  local circular speed, giving slightly different values. 
They found $v_\odot \ ({\rm km.s^{-1}})$ equal to 
$254 \pm 16$ \cite{reid}, 
$200-280$ \cite{mcmillan}, 
$236\pm11$ \cite{bovy},
$221 \pm 18$ \cite{Koposov:2009hn} and 
$218 \pm 7$  \cite{Feast:1997sb}. It is worth noticing that the evaluation of the Sun's circular velocity depends crucially on the value of  the
distance of the Sun to the galactic centre ($R_0$) which is also poorly known :  
$R_0 = 8.4\pm 0.4 \ {\rm kpc}$ \cite{ghez} and $R_0 =8.33 \pm 0.35  \ {\rm kpc}$ \cite{gillessen}.  In the following, the local circular speed is treated as a nuisance parameter, considering the range 
$220 \pm 30 \ {\rm km.s^{-1}}$, see tab.~\ref{tab:prior}.\\

 \setlength{\tabcolsep}{0.1cm}
\renewcommand{\arraystretch}{1.4}
\begin{table}[t]
\begin{center}
\hspace*{-0.5cm}
\begin{tabular}{|c||c|}
\hline
 Nuisance parameters  &   Gaussian parametrization \\ \hline \hline  
 $\rho_0 \ {\rm [GeV/c^2/cm^{3}]}$  &  $0.3\pm0.1$\\ \hline  
 $v_{\odot} \ {\rm [km/s]}$ & $220 \pm 30$ \\ \hline 
 $\sigma_x \ {\rm [km/s]}$ & $220/\sqrt{2} \pm 20$  \\ \hline 
 $\sigma_y \ {\rm [km/s]}$ & $220/\sqrt{2} \pm 20$  \\ \hline  
 $\sigma_z \ {\rm [km/s]}$ & $220/\sqrt{2} \pm 20$  \\ \hline 
\end{tabular}
\caption{Gaussian parametrization (mean and standard deviation) of the different astrophysical nuisance parameters.}
\label{tab:prior}
\end{center}
\end{table}
\renewcommand{\arraystretch}{1.1}

Finally, uncertainties in the local WIMP velocity distribution must be accounted for. The effect  of 
halo modeling on exclusion limits and allowed regions has been studied~\cite{billard.identification,green.jcap0708}.
Indeed, recent  results from N-body simulations are in favor of triaxial Dark Matter haloes with anisotropic velocity distributions 
and potentially containing substructure as sub-haloes (clumps) and dark disk \cite{Ling:2009cn,nezri,Bruch:2008rx,Read:2008fh,Tissera}. 
Moreover, recent observations of Sagittarius stellar tidal stream have shown evidence for a triaxial Milky Way Dark Matter halo \cite{Law:2009yq}.  However, 
it is noteworthy that this result holds true at large radius (60 kpc) and N-body simulations have shown that there can be significant
variations of the axis ratios with radius \cite{Hayashi}. S.~H.~Hansen and B.~Moore  identified a universal relation between 
the radial density slope and the velocity anistropy leading to  $\beta \sim 0.1$ at the solar neighbourhood \cite{Hansen:2004qs}. 
Using a sample of  1700 solar neighbourhood halo subdwarfs from the Sloan
Digital Sky Survey \cite{Bramich}, M.~C.~Smith {\it et al.} have evaluated the halo velocity dispersion and found the anisotropy 
parameter $\beta$ to be $\sim 0.5$ \cite{Smith:2009kr}. Interestingly, the stellar halo exhibits no net rotation.\\
Even if it should be noticed that halo stars have {\it a priori} a different density profile from the dark
matter, these facts can be taken as hints in favor of an anisotropic Dark Matter velocity distribution. To account for this effect, 
we consider a multivariate Gaussian WIMP  velocity distribution, {\it i.e.}  anisotropic, but without substructures. It is   
parametrized by the velocity dispersions $\sigma_{x,y,z}$.  Effect of non-smooth halo model with substructures and/or streams will be adressed in a forthcoming paper.\\
The velocity anisotropy $\beta(r)$, is  defined as \cite{biney},
  \begin{equation}
  \beta(r) = 1 - \frac{\sigma^2_{y} + \sigma^2_{z}}{2\sigma^2_x}
  \label{eq:beta}
  \end{equation}
According to N-body simulations with or without baryons  \cite{nezri,vogelsberger,Kuhlen:2009vh,Moore:2001vq,Teyssier:2001cp}, 
the $\beta$ parameter at a radius $R_{\odot} \simeq 8$ kpc from the Galactic center spans the  range 
$0-0.4$ which is in favor of radial anisotropy. Indeed, such radial anisotropy is expected as the gravitational potential is mainly a function of the radius.
In the following, the 
velocity dispersions $\sigma_{x,y,z}$ are treated as nuisance parameters,  considering the range 
$220/\sqrt{2} \pm 20 \ {\rm km/s}$, see tab.~\ref{tab:prior}, which corresponds to $\beta = 0 \pm 0.25$. We used a null mean value of the velocity anisotropy for the sake of comparison
 with other experiments, but the large uncertainty considered enables us to account for anisotropic halo model when computing the discovery potential of directional
 experiments.
Moreover, as   shown in \cite{billard.identification}, the effect of an extremely anisotropic halo model with $\beta = 0.4$ 
can be handled with such parametrization of the WIMP velocity distribution, as long as it can be approximated by a 
multivariate Gaussian, without  introducing bias in the estimation of the WIMP properties (mass and cross section).\\

In this study, we have not used the escape velocity $v_{\rm esc}$ as an astrophysical nuisance parameter, since directional detection with low energy threshold is 
almost insensitive to this parameter. Indeed, in the case of a WIMP mass of 100 GeV/c$^2$ the minimal speed to produce a 5 keV Fluorine recoil is 130 km.s$^{-1}$, far below
the median value of \cite{Smith:2006ym} equal to 544 km/s. Thus, the impact of $v_{\rm esc}$ is relevant for heavy targets and low WIMP mass
 in the so-called threshold region, where the experiment is only sensitive to the tail of the WIMP velocity distribution. In the following, we have considered an escape velocity
taken as infinity to simplify the calculations.\\

As a conclusion, the use of a profile likelihood ratio test statistic allows us to account for uncertainties on the 
 astrophysical parameters, as they are treated as nuisance parameters. This approach is a step beyond 
the "standard Dark Matter halo paradigm", {\it i.e.} isotropic isothermal Dark Matter halo with 
fixed value of density. Evaluating the properties of  the Dark Matter halo is indeed still 
a subject of debates and directional detection could bring valuable information, such as an evaluation of the anisotropy 
parameter, as shown in \cite{billard.identification}.

\section{The Profile Likelihood Ratio test statistic}
\label{sec:profile} 
 We are interested in estimating  the expected significance of a discovery of Dark Matter with directional detection. In this section, we  
 first introduce the definition of the likelihood function that we have used. In particular, astrophysical uncertainties are accounted for. Then, the formalism
 of a frequentist approach to estimate the significance using a profile likelihood ratio test statistic is presented. Finally, the impact of astrophysical
 uncertainties in the estimation of the significance is presented in a benchmark case.

\subsection{The likelihood function}
When considering directional detection of Dark Matter, the event distribution is given by the double differential spectrum (eq.~\ref{directionalrate}) which is a function of
both the recoil energy and direction. On the other hand, the background event distribution is expected to be 
isotropic and will be considered as flat in energy unless otherwise stated. In this study, we are also taking into account
five different astrophysical uncertainties as: the local Dark Matter density $\rho_0$, the circular velocity of the Sun $v_{\odot}$ and the velocity dispersion along 
the three axes of the halo $\{\sigma_x, \sigma_y,\sigma_z\}$. Then, considering the extended definition of the likelihood function (see Ref.\cite{Cowan}) to take into account
the Poisson statistics of the number of observed events $N$ and the
likelihood terms associated with each nuisance parameters, the full likelihood function is defined as,
\begin{align}
\mathscr{L}(\sigma_p,R_b,\vec{\nu}) & = \frac{(\mu_s + \mu_b)^N}{N!}e^{-(\mu_s + \mu_b)}\ \nonumber \\
 & \times\prod_{n = 1}^{N} \left[ \frac{\mu_s }{\mu_s + \mu_b} S(\vec{R}_n)  + \frac{\mu_b }{\mu_s + \mu_b}B(\vec{R}_n)\right ]\ \nonumber \\
 & \times\prod_{i = 1}^{5}\mathscr{L}_i(\nu_i)
 \label{eq:likelihood}
 \end{align}
where $\vec{\nu} = \{\rho_0,v_{\odot},\sigma_x, \sigma_y,\sigma_z \}$ represents the set of nuisance parameters, $\mu_b = R_b\times\xi$ and $\mu_s$ 
correspond to the number of expected background and WIMP events respectively, where $\xi$ corresponds to the exposure.
 $N$ is the total number of observed events, $\vec{R}_n$
 refers to the  energy and  
direction  of each event while the 
functions $S$ (see eq.\ref{directionalrate}) and $B$ (isotropic distribution for the angular part and following eq.\ref{eq:backmodel} for the energy part) 
are the directional event rate $d^2R/dE_Rd\Omega_R$ of the WIMP  and background 
  events respectively. The functions $\mathscr{L}_i(\nu_i)$ refer to the likelihood function associated with each astrophysical nuisance parameter. They are
 considered as Gaussian functions with the parametrization from Table~\ref{tab:prior}.

\subsection{The profile likelihood ratio}
The significance of a new process, in a frequentist approach, is commonly estimated by using the profile likelihood ratio test. 
We present hereafter a brief overview of the subject by  recalling basic relations for the reader's convenience. We refer to \cite{Cowan:2010js} for a comprehensive discussion.
The  profile likelihood ratio test corresponds to a
hypothesis test against the null hypothesis $H_0$ (background only) against the alternative $H_1$ which includes both background and signal. The strength of the profile
likelihood ratio relies in the fact that one can add some nuisance parameters both from an experimental side (efficiencies, resolutions,...) and from a theoretical side (signal or background
modeling). In our case, the $H_0$ hypothesis corresponds to the $\sigma_p = 0$ case (background only), while the alternative $H_1$ corresponds to the case where
$\sigma_p$ is taken different from 0 (background and signal). Then, in the case of a discovery, we test the background only hypothesis on the observed data and try to reject it using the following
ratio
\begin{equation}
\lambda(0) = \frac{\mathscr{L}(\sigma_p = 0,\hat{\hat{R_b}},\hat{\hat{\vec{\nu}}})}{\mathscr{L}(\hat{\sigma_p},\hat{R_b},\hat{\vec{\nu}})}.
\end{equation}
As discussed in \cite{Cowan:2010js} the test statistic in such case is defined as,

\begin{equation}
\rm q_0 = \left\{
\begin{array}{rrll}
\rm & -2\ln\lambda(0)	&	\ \hat{\sigma_p} > 0 \\
\rm & 0  		& 	\ \hat{\sigma_p} < 0
\end{array}\right.
\end{equation}

\begin{figure}[t]
\begin{center}
\includegraphics[scale=0.4,angle=0]{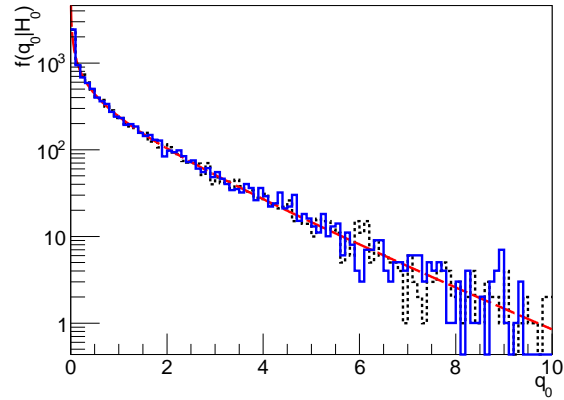}
\caption{Probability density functions of $q_0$ under the background only hypothesis $H_0$ with/without (blue solid line/black dotted line) 
considering astrophysical uncertainties, estimated with 10,000 Monte Carlo simulations. The red long dashed line correspond to the $\chi^2_1$ distribution.} 
\label{fig:Q0Check}
\end{center}
\end{figure}

As it can be seen from the previous equations, $0\leq\lambda(0)\leq1$ and $q_0 \geq 0$. Large value of $q_0$ implies a large discrepancy 
between the two hypotheses, which is in favor of a discovery. The $p$-value $p_0$ is defined as
\begin{equation}
p_0 = \int_{q^{\rm obs}}^{\infty}f(q_0|H_0)dq_0
\end{equation}
where $f(q_0|H_0)$ is the probability density function of $q_0$ under the background only hypothesis $H_0$. Then, $p_0$ corresponds to the probability to have a 
discrepancy, between $H_0$ and $H_1$, larger or equal to the observed one  $q^{\rm obs}$. As an example, a $p$-value of $p_0 = 0.00135$ corresponds to a 3$\sigma$ signal observation.
The main issue here is to have a correct estimation of the $f(q_0|H_0)$ distribution in order to compute the significance for a given data set. However, following Wilk's
theorem, $q_0$ asymptotically follows a $\chi^2$ distribution with one degree of freedom, see. \cite{Cowan:2010js}. Then, in such case, the discovery significance $Z$ 
is simply defined as $Z = \sqrt{q^{\rm obs}}$, in units of $\sigma$, {\it i.e.} $Z = 1$ corresponds to a significance of 68\%. 
Figure \ref{fig:Q0Check} presents the probability density functions $f(q_0|H_0)$ of $q_0$ under the background only hypothesis $H_0$ with/without 
astrophysical uncertainties. It is worth noticing that $f(q_0|H_0)$ is very well fitted by the $\chi^2_1$ distribution. Indeed, in both
cases: with and without astrophysical uncertainties, corresponding to the blue and black histograms, we found $\chi^2/{\rm d.o.f}$ equal to 95.27/99 and 99.79/99
respectively.

\subsection{Impact of astrophysical uncertainties}
\begin{figure}[t]
\begin{center}
\includegraphics[scale=0.4,angle=0]{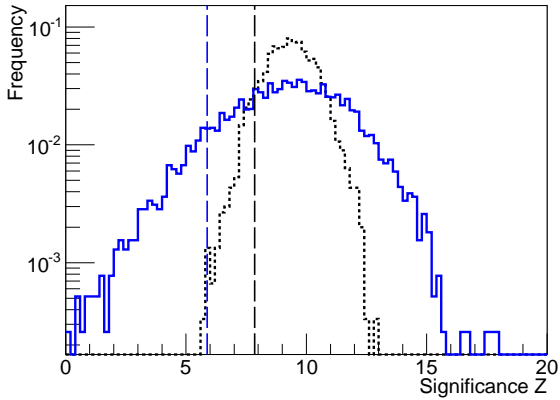}
\caption{Normalized distributions of the significance $f(Z)$ with/without (blue solid line/black dotted line) considering astrophysical uncertainties. Distributions were
generated using 10,000 Monte Carlo simulations with 100 WIMP events and 50 background events with a WIMP mass $m_{\chi} = 50$ GeV/c$^2$. We found mean values
of the significance of 9.2 and 9.3 and a 90\% confidence lower limit $Z_{90}$ of 7.9 and 5.9 respectively.} 
\label{fig:effect.astro}
\end{center}
\end{figure}

We aim at evaluating the discovery potential of upcoming directional detection experiment as a function of their experimental performance 
(resolutions, threshold, sense recognition) taking into account astrophysical uncertainties (see sec.~\ref{sec:astro.def}). In the following, we estimate the impact of the latter by studying
two different cases: with and without uncertainties from astrophysics. Not considering the latter corresponds to remove the $\mathscr{L}_i$ functions from 
eq.~(\ref{eq:likelihood}) and to
consider all the astrophysical parameters as perfectly known (without error bars). To do the comparison, we ran 10,000 Monte Calo simulations of a benchmark model with 100
WIMP events  and 50 background events with a WIMP mass of 50 GeV/c$^2$. The distributions of the significance $Z$ corresponding to the two scenarios are
presented on figure \ref{fig:effect.astro}. The black dashed histogram corresponds to the case without uncertainties and the blue solid one corresponds to the case where the
uncertainties are taken into account. One can easily deduce that the mean value of the significance E(Z) is almost the same, {\it i.e.} 9.3 and 9.2 respectively.
However, as shown on figure \ref{fig:effect.astro}, the spread of the $Z$  distribution is much wider when taking astrophysical uncertainties into account. 
This leads to  a large difference between the two cases when considering a confidence level on the value of $Z$.
Indeed, we defined $Z_{90}$ as the value of the significance that could be at least reached by an experiment 90\% of the time. Then $Z_{90}$ is found by solving the following
equation,
\begin{equation}
\int_0^{Z_{90}}f(Z)dZ = 0.9
\end{equation}
where $f(Z)$ corresponds to the normalized distribution of the significance Z. Then, the value of $Z_{90}$ with/without taking into account the astrophysical uncertainties is
5.9 and 7.9 respectively, leading to a large effect on the expected sensitivity to Dark Matter of a directional detector.\\
As a conclusion, this highlights the fact that it is necessary to take into account astrophysical uncertainties when estimating the sensitivity of a given directional
detection experiment to Dark Matter. In the following, we define the sensitivity of a given directional experiment by the lower bound of the 3$\sigma$ discovery region at 90\%
Confidence Level, given by the $Z_{90} = 3$ limit in the $(m_{\chi},\log_{10}(\sigma_p))$ plane.

\section{Optimizing directional detection}
\label{sec:effect}
All current directional projects \cite{drift,dmtpc,d3,emulsions,mimac,newage} face common challenges. As a matter of fact, experimental key issues include : sense recognition~\cite{dujmic,burgos,majewski}, 
angular and energy resolutions, energy threshold  as well as the 
residual background contamination.\\ 
In this section, we evaluate the evolution of the significance of a discovery as a function of the detector performance. In a first place, we will be interested 
in the effect of the number of WIMP and background events in the data. Then, in a second part, we will consider some 
experimental issues, like the energy threshold $E_{\rm th}$, the angular resolution $\sigma_{\gamma}$, the sense recognition efficiency, the energy resolution $\sigma_{E_r}$ and the energy
background modeling.
In the following, we   consider the mean significance $\rm E(Z)$ and the sensitivity, estimated from the limit corresponding to $Z_{90} = 3$ in the
$(m_{\chi},\log_{10}(\sigma_p))$ plane, using 1,000 Monte Carlo simulations for each given set of input parameters. 
Unless otherwise stated, we   consider a 10 kg CF$_4$
directional detector with a recoil energy range of [5, 50] keV with perfect angular and energy resolutions and with sense recognition.
To study one by one the impact of a given experimental issue on the sensitivity, this parameter will be degraded keeping all other ones
unchanged. 

\subsection{Residual background contamination} 
Zero background is often referred to as the ultimate 
goal for the next generation of direct detection experiments in 
deep underground laboratories. However, owing to the large intrinsic difference between the WIMP-induced and
background-induced  spectra, directional detection could accommodate to a sizeable background 
contamination. In this section, we investigate the effect of a residual background contamination, {\it i.e.} in the final dataset after all selections based {\it e.g.} on
track-length versus energy and fiducialization. Hence, we try to account for the effect of an irreducible background contamination, which can only be treated by means of statistical analysis.\\

\begin{figure}[t]
\begin{center}
\includegraphics[scale=0.4,angle=0]{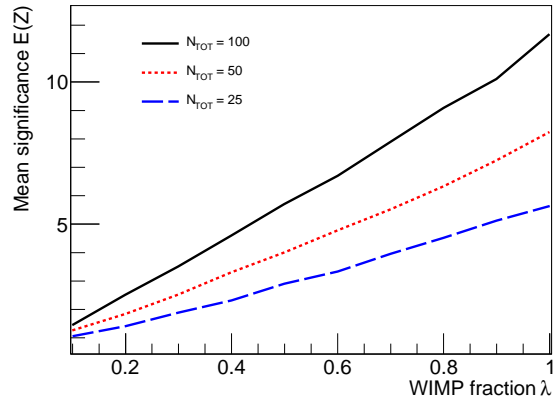}
\caption{Evolution of the mean significance E(Z) as a function of the expected WIMP fraction $\lambda = N_s/(N_s + N_b)$ 
for three different values of the expected total number of events $N_{tot} = 100$ 
(black solid line), 50 (red dotted line) and 25 (blue long dashed line). This study has been done by considering a WIMP mass of 50 GeV/c$^2$.} 
\label{fig:Lambda}
\end{center}
\end{figure}

To begin with, on figure \ref{fig:Lambda} we studied the effect of background/signal contribution to data. Figure \ref{fig:Lambda} represents the evolution of
 the mean significance E(Z) as a function of $\lambda = N_s/(N_s+N_b)$, which corresponds to the fraction of WIMP events contained in the data, for 
 three different total numbers of events $N_{tot} = 25, 50, 100$. Hence, $\lambda \rightarrow 0$
 means that the data are background dominated.
Then, from figure \ref{fig:Lambda}, one can observe that for any total number of events, the mean significance E(Z) increases almost linearly with the data purity ($\lambda$). Also, for a
 given value of $\lambda$, increasing the total number of events will enhance the significance. In other words, for a given value of $\lambda = 0.5$ for example,
the significance of a Dark Matter detection can be improved from 3$\sigma$ to 6$\sigma$ by having an exposure 4 times larger.\\

\begin{figure}[t]
\begin{center}
\includegraphics[scale=0.4,angle=0]{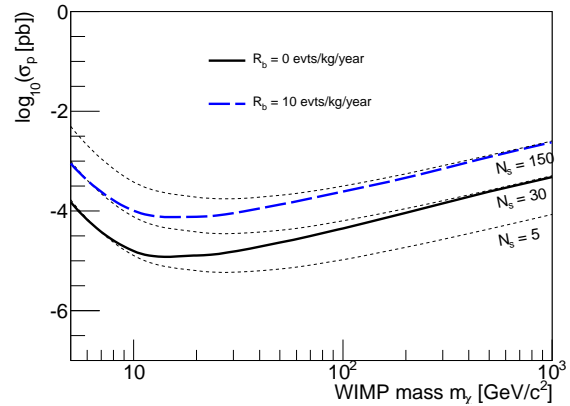}
\caption{Lower bound of the 3$\sigma$ discovery region at 90\% C.L. in the ($m_\chi,\log_{10}(\sigma_p)$) plane. Black line presents  the
background free case, while the blue dashed  line
presents the same region with 10 background events per year per kg. For convenience,  the curves of iso-number of WIMP events are presented (dashed lines) in the case of 
5, 30 and 150 WIMP events.} 
\label{fig:BackgroundImpact}
\end{center}
\end{figure}

Figure  \ref{fig:BackgroundImpact} presents   the lower bound of the 3$\sigma$ discovery region at 90\% C.L. ($Z_{90} = 3$ limit) 
in the ($m_\chi,\log_{10}(\sigma_p)$) plane. Two cases are presented :  background-free  and high background 
contamination ($10 \ {\rm kg^{-1}.year^{-1}}$). For convenience, the curves of iso-number of WIMP events are presented (dashed lines) in the case of 5, 30 and 150 WIMP events. 
One can first notice that a 30 kg.year $\rm CF_4$ directional detector may allow to achieve, in the background-free case,  a 3$\sigma$ discovery of Dark Matter down to $\sim 10^{-5}$ pb at 
low WIMP mass ($\sim 10 \ {\rm GeV/c^2}$) and 
down to $\sim 10^{-4}$ pb for high WIMP mass ($\sim 1 \ {\rm TeV/c^2}$). Adding a large fraction of background in the data ($10 \ {\rm kg^{-1}.year^{-1}}$) results in a loss 
of about one order of magnitude in the lower bound of the $3\sigma$ discovery region. This highlights the fact that directional detection can accommodate to a sizeable background 
contamination, noticing that such a background event rate in the final dataset is very large. Interestingly, one may compare the 3$\sigma$ lower bound to 
 the curves of iso-number of WIMP events.  At low WIMP mass ($\sim 10 \ {\rm GeV/c^2}$) about 5 WIMP events are enough to claim a $3\sigma$ discovery, in the background-free
  case, while about 30 WIMP events are needed for
 highly-background contaminated data. At high WIMP mass  ($\sim 1 \ {\rm TeV/c^2}$) these numbers reach 30 and 150 respectively.\\
To conclude on this first study, aiming at evaluating the effect of $N_s$ and $N_b$ on the mean significance, one can easily appreciate the fact that using directional
 detection, strong evidence in favor of a Dark Matter detection could be reached even with low statistics and rather large background contamination. 
This low sensitivity to background contamination and the possibility to evaluate the Dark Matter and background components in the final 
dataset,  give a major interest for directional detection, especially at 
the present stage when non-directional experiments start to observe candidate events 
whose origin is difficult to assess 
\cite{Ahmed:2011gh,Aprile:2011hi,Aalseth:2010vx,Angloher:2011uu}. On the other hand, we emphasize that, for directional detection, an
 unambiguous proof of discovery of Galactic Dark Matter would be given
by a main recoil direction pointing toward the constellation Cygnus within few degrees, combined with a high significance. However, we caution that the presence of a stream
or another Dark Matter substructure in the vicinity of the Solar System could deviate the recovering angle from the direction 
of the constellation Cygnus of a few degrees, depending on the relative densities, as outlined in \cite{Morgan:2004ys}.

\subsection{Effect of the energy threshold}
As for direction-insensitive direct detection, the energy threshold plays a key role for directional detection. 
It is worth emphasizing that it is the lowest energy at which both the initial direction and the energy of the recoiling nucleus can be retrieved, 
which makes it
even more challenging for directional detection. In particular, this ``directional threshold'' is higher than the standard 
energy threshold  defined as the minimal 
kinetic energy required for an event to be detected.
Indeed, a low energy recoil in a low pressure TPC would present a short track length and a 
   large angular dispersion, implying a  loss of the direction information. 
   The directional energy threshold is thus closely related to the gas pressure, the target choice, the readout performance as well as the data analysis strategy.
There are two main and competing effects when increasing the energy threshold : a reduction of the number of  
expected WIMP events and a selection of the most anisotropic WIMP-induced recoils.\\

\begin{figure}[t]
\begin{center}
 \includegraphics[scale=0.4,angle=0]{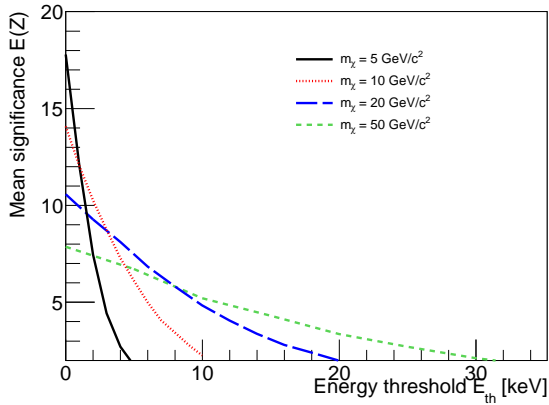}
\caption{Mean significance E(Z) as a function of the energy threshold, for four different WIMP masses : 
$m_{\chi} = $ 5  (black solid line), 10  
(red dotted line), 20   (blue long dashed line) and 50 GeV/c$^2$ (green short dashed line). We have considered a constant background rate of
$R_b = 10$ evts/kg/year and different values of $\sigma_p$ to get $\mu_s = 100$ events for each WIMP mass at $E_{th} = 0$ keV.} 
\label{fig:Threshold}
 \end{center}
\end{figure}

On figure~\ref{fig:Threshold}, we present the evolution of $E(Z)$ as a function of $E_{th}$ for four different WIMP masses : 
$m_{\chi} = $ 5  (black solid line), 10  
(red dotted line), 20   (blue long dashed line) and 50 GeV/c$^2$ (green short dashed line). 
For the sake of comparison between the different WIMP masses, we have chosen
cross section values in order to have 100 WIMP events in each case at $E_{th} = 0$ keV and a constant background contamination of 10 evts/kg/year. 
As discussed in \cite{billard.disco}, the fact that the
angular recoil distribution of WIMP events is more anisotropic at low WIMP mass comes from the fact that the energy threshold selects events at higher energies
 which are the most anisotropic ones. 
The fact that, at $E_{th} = 0$ keV, the significance is higher for lighter WIMPs is only due to the energy spectrum which is more peaked at low energy and then
more different to the background one, assumed to be flat. Then, increasing the energy threshold will decrease the number of WIMP events more rapidly in the case of light
WIMPs than for heavier WIMPs and also enhance the anisotropy related to the rejection power.
 However, as it can be seen from figure \ref{fig:Threshold}, the significance is always decreasing  with the energy threshold leading to the conclusion that the loss of WIMP
 events has a stronger effect on the significance than the enhancement of the anisotropy. 
 Then, contrarily to what was predicted by \cite{Green:2006cb}, when looking at the whole directional event distribution $d^2R/dE_rd\Omega_r$, 
 there is no enhancement of the
 significance at low energy threshold. Of course, we caution that our result presented here suffers from energy background 
 modeling dependence, 
 and that for other  assumptions, one could find a result similar to the one found in \cite{Green:2006cb} which as been 
 obtained by considering only the angular distribution of the
 events.\\

 \begin{figure}[t]
\begin{center}
\includegraphics[scale=0.4,angle=0]{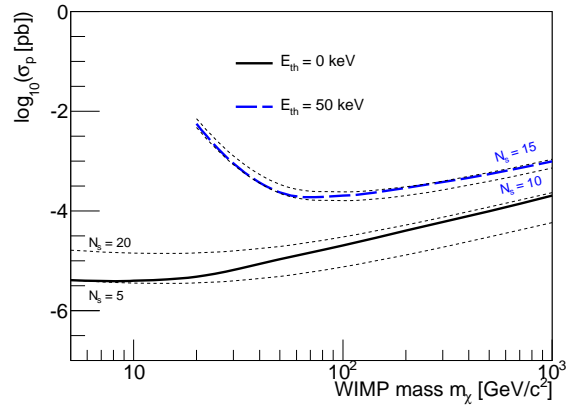}
\caption{Lower bound of the 3$\sigma$ discovery region at 90\% C.L. in the ($m_\chi,\log_{10}(\sigma_p)$) plane for two different cases: $E_{th} = 0$ keV (black solid line) and $E_{th} = 50$
keV (blue long dashed line) without background events. For convenience,  the curves of iso-number of WIMP events are presented in the $E_{th} = 0$ keV  case (black dashed lines) and in the 
$E_{th} = 50$ keV case (blue dashed lines).} 
\label{fig:ThresholdImpact}
\end{center}
\end{figure} 

Figure \ref{fig:ThresholdImpact} presents  
the  
lower bound of the 3$\sigma$ discovery region at 90\% C.L. in the ($m_\chi,\log_{10}(\sigma_p)$) for two different 
cases : $E_{th} = 0$ keV  (black solid line) and $E_{th} = 50$ keV (blue long dashed line).  
For convenience, the curves of iso-number of WIMP events, relevant to each case, are  presented (dashed lines).  
One can first notice that a 30 kg.year $\rm CF_4$ directional detection would allow to 
achieve, in the unrealistic $E_{th} = 0$ keV case,  a 3$\sigma$ discovery of Dark Matter down to $\sim 10^{-5}$ pb at 
low WIMP mass ($\sim 10 \ {\rm GeV/c^2}$) and down to $\sim 10^{-4}$ pb for high WIMP mass ($\sim 1 \ {\rm TeV/c^2}$). At high
WIMP mass, about 4 times more WIMP events are required to achieve a 3$\sigma$ discovery.  
Considering a detector with a 50 keV energy threshold  results in a loss 
  of a factor of 5 in sensitivity at high WIMP mass and to no sensitivity at all below $\sim$ 20 GeV/c$^2$.\\
However, as one can see from figure \ref{fig:ThresholdImpact}, the fact that, at high WIMP mass,
 only 15 WIMP events are required to get a 3$\sigma$ discovery at 90\% C.L. for  $E_{th} = 50$ keV
while 20 are required for $E_{th} = 0$ keV comes from the enhancement of the anisotropy due to the high energy threshold.\\

 As a conclusion of this study aiming at quantifying the impact of the energy threshold on the expected sensitivity 
 of a given directional detection, it can simply be stated that the lower remains as expected the better. 
 We also found that a low energy threshold is compulsory to be sensitive to low WIMP mass. 
 As an illustration of this statement, we found that a threshold of about 50 keV will prevent a 
 given directional experiment, using Fluorine target material, to be sensitive to very low WIMP mass $\sim 5 - 10$ GeV/c$^2$. 
 We conclude that the energy threshold remains a major experimental issue for directional detection, noticing 
 that both the 3D track and the energy must be measured down to this energy.


\subsection{Effect of the sense recognition efficiency}
Not only should the track be
3D-reconstructed, but its sense should also be retrieved from the data analysis. 
As outlined in \cite{Billard.cygnus}, an asymmetry between upgoing and downgoing tracks is expected, due to 
two different effects. First, the angular dispersion of recoiling tracks should result in a spatial asymmetry as 
the beginning of the track should be more rectilinear than its end. Second, a charge collection asymmetry is expected as the $dE/dx$ in ionization is 
decreasing with energy at low recoil energy. Hence, more primary electrons should be generated at the beginning of the track.\\
Even though several experimental progresses have been done \cite{dujmic,burgos,majewski}, sense recognition 
remains a key and challenging experimental issue for directional detection of Dark Matter. In particular, it 
should still be demonstrated that sense recognition may be achieved at low recoil energy, where most WIMP events reside.\\ 

 \begin{figure}[t]
\begin{center}
\includegraphics[scale=0.4,angle=0]{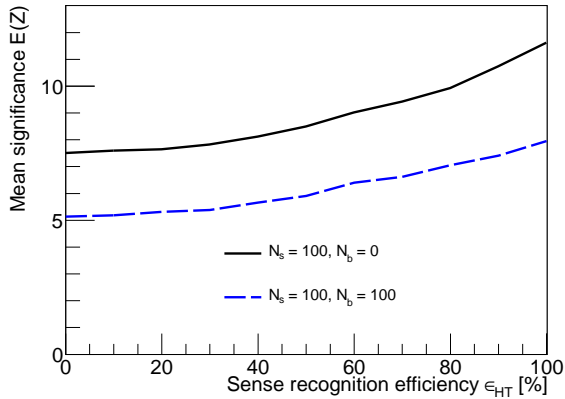}
\caption{Mean significance as a function of $\epsilon_{HT}$ for two different cases ($N_s = 100, N_b = 0$) (black solid line) and 
($N_s = 100, N_b = 100$) (blue long dashed line). A WIMP mass of 50 GeV/$c^2$ has been considered for this study. } 
\label{fig:HeadTail}
\end{center}
\end{figure} 

In the following,  
we investigate the effect  of no or partial sense recognition on the expected significance of a given directional detector. 
To do so, we define a sense recognition efficiency   as :
$\epsilon_{HT} = \epsilon_G - \epsilon_W$, where $\epsilon_G$ and $\epsilon_W$ corresponds to the fraction of good and 
wrong reconstructed events respectively. Then, if 
$\epsilon_{HT} = 0 \%$, we are in the ``flipping coin'' 
scenario, {\it i.e.} the detector does not have any   sense recognition capability. 
Then, the modified  directional event rate that takes into account this partial sense recognition $S(\hat{R}_n)$ is defined as :
\begin{equation}
S(\hat{R}_n) = \frac{1+\epsilon_{HT}}{2}S(+\hat{R}_n) + \frac{1-\epsilon_{HT}}{2}S(-\hat{R}_n),
\end{equation}
where $\hat{R}_n$ refers to the direction of the recoiling nucleus. \\

 \begin{figure}[t]
\begin{center}
\includegraphics[scale=0.4,angle=0]{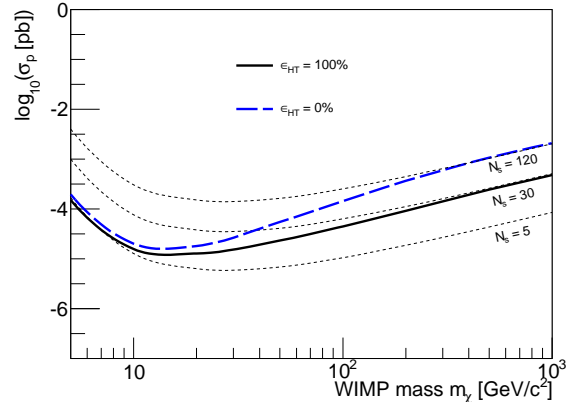}
\caption{Lower bound of the 3$\sigma$ discovery region at 90\% C.L. in the ($m_\chi,\log_{10}(\sigma_p)$) plane for two different cases: $\epsilon_{HT} = 100$\% 
(black solid line) and $\epsilon_{HT} = 0$\% (blue long dashed line) without background events. 
For convenience,  the curves of iso-number of WIMP events are presented (dashed lines) in the case of $N_s$ = 5, 30 and 120.}  
\label{fig:ProspectHeadTail}
\end{center}
\end{figure} 

Without sense recognition, the expected WIMP-induced distribution becomes less anisotropic 
and thus gets closer to the expected background event distribution. This induces an obvious loss of discrimination power.
Figure \ref{fig:HeadTail} presents the evolution of the mean significance E(Z) as a function of 
$\epsilon_{HT}$ for different cases ($N_s = 100, N_b = 0$) and 
($N_s = 100, N_b = 100$) considering a flat energy spectrum for the background.
 One can notice that, for both cases, the mean significance is monotonically increasing with $\epsilon_{HT}$. 
It is also shown that this significance improvement   is even stronger for values of $\epsilon_{HT}$ close to 100\%. It 
means that having $\epsilon_{HT}$ equal to 0\% or to 20\% does not really modify the expected
significance while having  $\epsilon_{HT}$ equal to 80\% or to 100\% changes the mean significance by almost 
20\%. Also, it is shown on 
figure \ref{fig:HeadTail} that even in the presence of a sizeable background contamination ($\lambda = 0.5$)
 the expected mean significance remains strong, {\it i.e.} above 5$\sigma$. 
 Indeed, the presence of background reduces the mean significance by about 30\% all over the range of $\epsilon_{HT}$.\\

Figure \ref{fig:ProspectHeadTail} presents 
the  
lower bound of the 3$\sigma$ discovery region at 90\% C.L. in the ($m_\chi,\log_{10}(\sigma_p)$) plane for two different 
cases: $\epsilon_{HT} = $100\% (black solid line) and $\epsilon_{HT} = $0\% (blue long dashed line).  For convenience, the curves of iso-number of WIMP events are 
presented (dashed lines) in the case of 5, 30 and 120 WIMP events. 
One can first notice that a 30 kg.year $\rm CF_4$ directional detection would allow to 
achieve, with full sense recognition capability,  a 3$\sigma$ discovery of Dark Matter down to $\sim 10^{-5}$ pb at 
low WIMP mass ($\sim 10 \ {\rm GeV/c^2}$) and down to $\sim 10^{-4}-10^{-3}$ pb for high WIMP mass ($\sim 0.1-1 \ {\rm TeV/c^2}$). At high
WIMP mass, about 6 times more WIMP events are required to achieve a 3$\sigma$ discovery.  
Considering a detector with no sense-recognition  capability results in a loss 
of a factor of 4 at high WIMP mass and almost no effect at low WIMP mass. Note that, this statement depends  on
the background energy distribution (see sec.~\ref{sec:bckmodel}).\\
This study leads us to the conclusion that sense recognition is not a major experimental issue. Indeed, 
a directional detector  without sense recognition capability would still be able to achieve a 3$\sigma$ discovery in the 
$10^{-5}-10^{-3} \ {\rm pb}$ region. We emphasize that the same conclusion holds true when trying to set exclusion limits, 
as shown in \cite{billard.exclusion}. This result is of major interest,  as getting sense recognition from experimental data 
remains a very difficult task, probably the most difficult experimental challenge to be faced by current projects, 
and especially when trying to get a high sense recognition efficiency down to the energy threshold.

\subsection{Effect of angular resolution}

 \begin{figure}[t]
\begin{center}
\includegraphics[scale=0.4,angle=0]{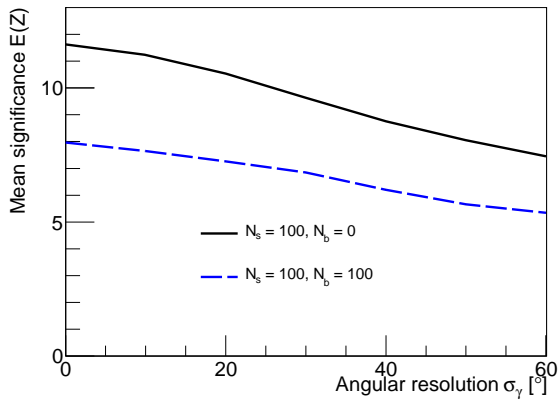}
\caption{Mean significance as a function of the angular resolution 
$\sigma_{\gamma}$ for different cases ($N_s = 100, N_b = 0$) (black solid line) and ($N_s = 100, N_b = 0$) (blue long dashed line). A WIMP mass of 50 GeV/c$^2$ has been
considered for this study.} 
\label{fig:AngularResolution}
\end{center}
\end{figure} 
As far as directional detection is concerned, the estimation of the initial recoil direction is compulsory.  This gives an intrinsic limitation of this detection 
strategy as recoil  tracks in low
 pressure gaseous detectors would encounter a rather large angular dispersion ("straggling" effect) by colliding with other nuclei of the gas.
  Moreover, the gas
 properties imply  a transverse and longitudinal diffusion of the primary electrons that will contribute to the angular resolution.\\ 
 Hence,  data of upcoming directional detectors should suffer from rather large angular resolution. 
Having a finite angular resolution means that a recoil initially coming from the direction  $\hat{r}(\Omega_r)$ is 
reconstructed as a recoil coming from the direction ${\hat{r}}^{\, \prime}(\Omega_r^\prime)$ with a gaussian dispersion of 
 width $\sigma_\gamma$ according to the following distribution:
\begin{equation}
K(\Omega_r,\Omega'_r) = e^{-\gamma^2/2\sigma_{\gamma}^2}/\left((2\pi)^{2/3}\sigma_{\gamma}{\rm erf}(\sqrt{2}\sigma_{\gamma})\right)
\end{equation}
where the angle $\gamma$ between the $\hat{r}(\Omega_r)$ and ${\hat{r}}^{\, \prime}(\Omega_r^\prime)$ is defined by
\begin{equation}
\cos{\gamma} = \cos b_r\cos b_r'\cos (l_r-l_r') + \sin b_r \sin b_r',
\end{equation}
where the angles $l$ and $b$ refer  to the Galactic coordinates.\\
The directional recoil rate is then given by the convolution product of the initial recoil rate and the angular ditribution as
\begin{equation}
\frac{d^2R}{d\Omega_r dE_r}(\Omega_r,E_r) = \int_{\Omega_r'}\frac{d^2R}{d\Omega_r dE_r}(\Omega_r',E_r) K(\Omega_r,\Omega_r')d\Omega_r'.
\end{equation}

 \begin{figure}[t]
\begin{center}
\includegraphics[scale=0.4,angle=0]{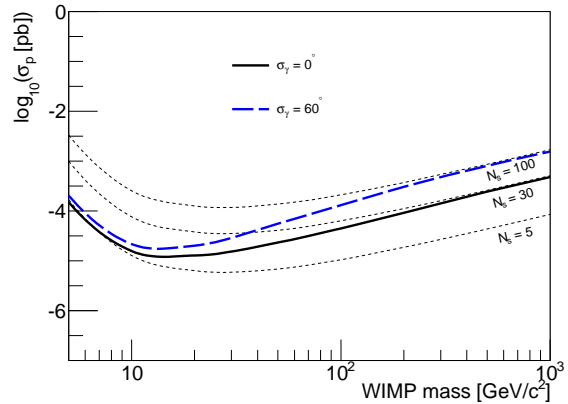}
\caption{Lower bound of the 3$\sigma$ discovery region at 90\% C.L. in the ($m_\chi,\log_{10}(\sigma_p)$) plane for two different cases: $\sigma_{\gamma} = 0^{\circ}$ 
(black solid line) and $\sigma_{\gamma} = 60^{\circ}$ (blue long dashed line) without background events.
For convenience,  the curves of iso-number of WIMP events are presented (dashed lines) in the case of 5, 30 and 100  WIMP events.} 
\label{fig:AngularProspect}
\end{center}
\end{figure}

The evolution of the expected mean significance E(Z) as a function of the angular resolution $\sigma_{\gamma}$ is shown on figure \ref{fig:AngularResolution}. Two different
cases are considered  $(N_s=100,N_b=0)$ (black solid line) and $(N_s=100,N_b=100)$ (blue long dashed line) with a flat background. 
As one can see,  for each configuration, degrading the angular resolution ({\it i.e.} increasing $\sigma_{\gamma}$) results in a 
diminution of the expected mean significance. This comes from the fact that for large angular
values of the angular resolution, the angular distribution of the expected WIMP events is less anisotropic and is getting closer to the angular 
distribution of background events, which is
isotropic. From figure \ref{fig:AngularResolution}, we can see that an angular resolution of $\sigma_{\gamma} = 60^{\circ}$ decreases the mean significance of about 30\% 
in comparison to the case where $\sigma_{\gamma} = 0^{\circ}$. \\

The effect of angular resolution on the expected discovery potential of a directional detection experiment is shown on figure \ref{fig:AngularProspect}. Indeed, we have 
represented in the ($\log_{10}(\sigma_p), m_{\chi}$) plane, the 90\% C.L. limit to reach a Dark Matter detection with a significance greater or equal to 3$\sigma$ considering
a perfect angular resolution (black solid line) and $\sigma_{\gamma} = 60^{\circ}$ (blue long dashed line). For convenience, the black dashed line corresponds to iso-value of
expected WIMP events for $N_s$ = 5, 30 and 100. As one can see, the effect of an angular resolution of 60$^{\circ}$ is very small at low WIMP mass due to the fact that at low
WIMP mass the expected angular distribution is much more anisotropic than for heavy WIMPs. Then, at high WIMP mass, the effect of angular resolution is non negligible as we
found a reduction in the sensitivity of about a factor of 3.3. This result suggests that directional detection should still be very competitive even with a rather poor angular
resolution. It may be noticed from figure \ref{fig:ProspectHeadTail} and figure \ref{fig:AngularProspect} that the effect of an angular resolution of
60$^{\circ}$ with $\epsilon_{HT} = 100$\% is almost equivalent to not having the sense recognition capability with a perfect angular resolution. \\
Then, as for the sense recognition study, we can conclude that angular resolution  
is not a major experimental issue. Indeed, a low angular resolution detector  would still be able to achieve a 3$\sigma$ discovery in the 
$10^{-5}-10^{-3} \ {\rm pb}$ region.

\subsection{Effect of the energy resolution}
Low pressure gaseous TPC detectors allow the ionization energy to be
measured. It should then be converted to a recoil energy thanks to the knowledge of 
the ionization quenching factor. The measurement of this quantity is a major issue for directional detection with gaseous TPC \cite{santos.q,guillaudin.cygnus}. \\
So far, we have considered a directional detector with a perfect energy resolution. In this section, we present the effect of a 
non-perfect energy resolution on the expected
sensitivity of a directional detector. As outlined above, most directional detection experiments are TPCs, they will hence be affected by 
energy resolutions much higher than the ones from
cryogenic detectors for instance.\\
The energy resolution on the recoil energy $\sigma_{E_r}$ is taken into account in the expected directional recoil rate distribution using the following procedure:
\begin{align}
\frac{d^2R}{d\Omega_r dE_r}(\Omega_r,E_r)& =  \int_{E'_r}\frac{d^2R}{d\Omega_r dE_r}(\Omega_r,E'_r) \ \times  \\ \nonumber
&  \frac{1}{\sqrt{2\pi}\sigma(E'_r)}\exp{\left\{-\frac{1}{2}\left(\frac{E_r - E'_r}{\sigma(E'_r)}\right)^2\right\}}dE'_r.
\end{align}

 \begin{figure}[t]
\begin{center}
\includegraphics[scale=0.4,angle=0]{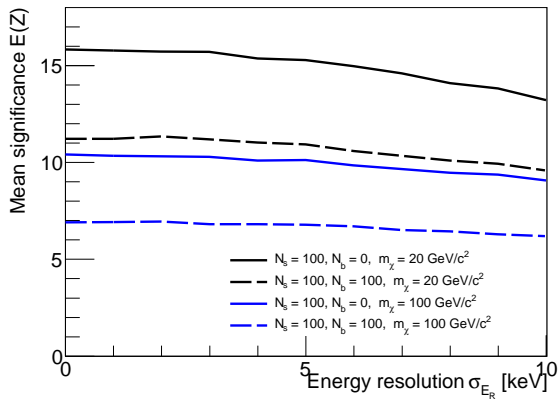}
\caption{Mean significance as a function of 
 $\sigma_{E_r}$ for different cases ($N_s = 100, N_b = 0$) (solid lines) and 
($N_s = 100, N_b = 100$) (long dashed lines) according to two different WIMP masses: $m_{\chi} = 20$ GeV/c$^2$ (black lines) and $m_{\chi} = 100$ GeV/c$^2$ (blue lines).} 
\label{fig:EnergyReso}
\end{center}
\end{figure}

Having a non-perfect energy resolution will lead to two different effects that are in competition. The first one is that it will 
enhance the expected number of WIMP events.
Indeed, as most of the WIMP events lie at low energy, if the energy resolution ($\sigma_{E_r}$) is large, low energy events could pass the energy threshold and be detected.
The second effect is that it will smooth the energy distribution leading to a WIMP event energy distribution closer to the expected background one (flat) and then reduce the
discrimination power.\\
The effect of energy resolution is shown on figure \ref{fig:EnergyReso} which presents the evolution of the mean expected significance E(Z) as a function of $\sigma_{E_r}$.
The latter presents the results obtained in the cases of a 20 GeV/c$^2$ (black lines) and 100 GeV/c$^2$ (blue lines) WIMP mass considering  $(N_s=100,N_b=0)$ (solid lines)
and $(N_s=100,N_b=100)$ (long dashed lines). We have considered value for $\sigma_{E_r}$ ranging from 0 keV, {\it i.e.} perfect energy resolution, to 10 keV. As the recoil
energy range of the detector is taken, for this example, between 5 and 50 keV, $\sigma_{E_r} = 10$ keV corresponds to an extreme and unrealistic energy resolution of 200\% at 5 keV 
and 20\% at 50 keV. However, as one can see from figure \ref{fig:EnergyReso}, even with an extremely large energy resolution, the mean expected significance is only slightly
degraded compared to the case of a perfect energy resolution, {\it i.e.} 15\% in the case $m_{\chi} = 20$ GeV/c$^2$ and $(N_s=100,N_b=0)$ (black solid line). From figure
\ref{fig:EnergyReso} we can also appreciate the fact that the effect of energy resolution on the significance of a Dark Matter detection is negligible even with a 50\% background
contribution and for a 100 GeV/c$^2$ WIMP mass.\\
As a conclusion of this study, we have shown that energy resolution is   a meaningless experimental issue for 
the Dark Matter sensitivity of a given directional detector.

\subsection{Effect of the background energy model}
\label{sec:bckmodel}
\begin{figure}[t]
\begin{center}
\includegraphics[scale=0.4,angle=0]{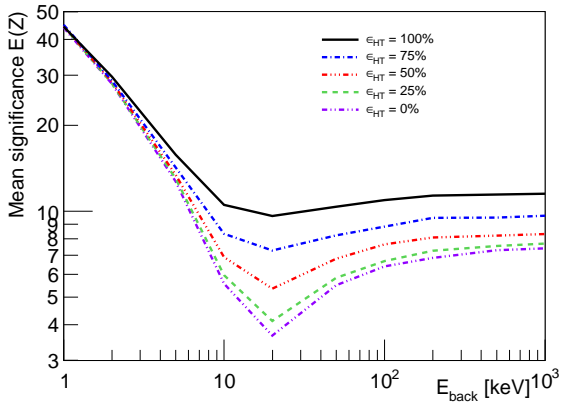}
\caption{Mean significance as a function of 
the background slope $E_{\rm back}$ for 5 different values of sense recognition efficiencies (from top to bottom): $\epsilon_{HT}$ = 100\%, 75\%, 50\%, 25\% and 0\%. A WIMP
mass of 50 GeV/c$^2$ with 100 expected WIMP events and no background contamination has been considered in this study.} 
\label{fig:BackSlope}
\end{center}
\end{figure}

\begin{figure}[t]
\begin{center}
\includegraphics[scale=0.4,angle=0]{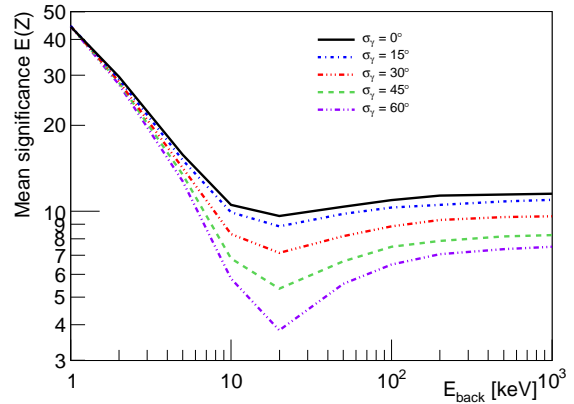}
\caption{Mean significance as a function of 
the background slope $E_{\rm back}$ for 5 different values of angular resolution (from top to bottom): $\sigma_{\gamma}$ = 0$^{\circ}$, 15$^{\circ}$,
 30$^{\circ}$, 45$^{\circ}$ and 60$^{\circ}$. A WIMP
mass of 50 GeV/c$^2$ with 100 expected WIMP events and no background contamination has been considered in this study.} 
\label{fig:BackSlopeAngular}
\end{center}
\end{figure}

The last experimental issue to be discussed  is the effect of energy background 
modeling on the estimation of the sensitivity of a directional detector.
In the previous sections, we have considered a directional event rate for background events flat in energy. Then, in this
section, we   consider an exponential background energy spectrum in the form of,
\begin{equation}
\left.\frac{dR}{dE_{r}}\right|_{\rm back} =  -\frac{1}{E_{\rm back}}\times\frac{\exp{(-{E_{r}}/{E_{\rm back}})}}{\exp{(-{E_{th}}/{E_{\rm back}})}}
\label{eq:backmodel}
\end{equation}
where $E_{\rm back}$ refers to the slope of the background energy distribution and if $E_{\rm back} \rightarrow +\infty$ we recover a 
flat energy spectrum. Obviously, the
worst  scenario is when the expected WIMP event energy distribution and the background one are the same. Indeed, in such case, there is 
no more discrimination between WIMP events and background events based on the energy information.
For concreteness, we examplify this statement by considering the case $E_{\rm back} = 17.4$ keV, corresponding to the slope 
expected for WIMP-induced recoil energy distribution when considering a WIMP mass of 50 GeV/c$^2$ and  Fluorine target. 
On figure \ref{fig:BackSlope} and \ref{fig:BackSlopeAngular}, one can see the evolution of E(Z) as a function of $E_{\rm back}$ for different values of $\epsilon_{HT}$ and
$\sigma_{\gamma}$. The common feature is that E(Z) is maximal at very low value of $E_{\rm back}$ corresponding to a very steep background energy spectrum, then decreases to a
minimal value located around $E_{\rm back} = 17.4$ keV and then increases asymptotically to the case where  $E_{\rm back} \rightarrow +\infty$ corresponding to a flat
banckground energy spectrum. An interesting point that should be highlighted is the fact that the effect of $E_{\rm back}$ on the expected sensitivity 
depends strongly on the angular performance of the detector. Indeed, the minimal value of E(Z) at $E_{\rm back} = 17.4$ keV is 
strongly degraded when decreasing the sense
recognition efficiency or increasing the angular resolution. 
This is simply explained by the fact that, for this particular value of the background slope exactly equal to the WIMP-induced  event 
slope, the discrimination between WIMP and background events relies only on the angular information.

\begin{figure}[t]
\begin{center}
\includegraphics[scale=0.4,angle=0]{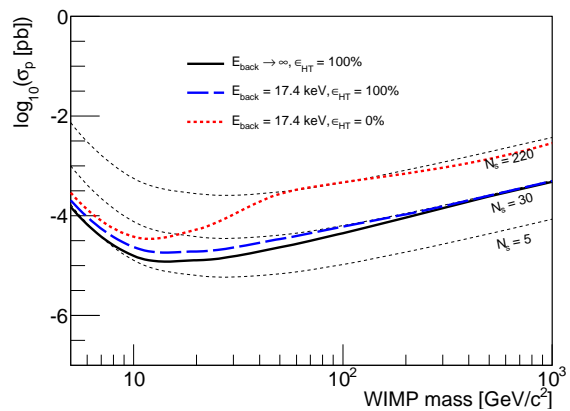}
\caption{Lower bound of the 3$\sigma$ discovery region at 90\% C.L. in the ($m_\chi,\log_{10}(\sigma_p)$) plane for three different cases: ($E_{\rm back} \rightarrow +\infty$ keV,
 $\epsilon_{HT} = 100$\%) (black solid line), ($E_{\rm back} = 17.4$ keV, $\epsilon_{HT} = 100$\%) (blue long dashed line) and
  ($E_{\rm back} = 17.4$ keV, $\epsilon_{HT} = 0$\%) (red dotted line) without background events. 
For convenience,  the curves of iso-number of WIMP events are presented (dashed lines) in the case of 5, 30 and 220  WIMP events.} 
\label{fig:BackSlopeProspect}
\end{center}
\end{figure}

Finally, the effect of energy background modeling on the expected discovery potential of a given directional detector is shown on figure \ref{fig:BackSlopeProspect}.
 We have
presented, on the ($m_{\chi}, \log_{10}(\sigma_p)$) plane, 
the  
lower bound of the 3$\sigma$ discovery region at 90\% C.L. in three different cases: flat background energy spectrum
($E_{\rm back} \rightarrow +\infty$) with $\epsilon_{HT} = 100$\% (black solid line), $E_{\rm back} = 17$ keV with $\epsilon_{HT} = 100$\% (blue long dashed line) and 
$E_{\rm back} = 17$ keV with $\epsilon_{HT} = 0$\% (red dotted line). Taken at face value, these results suggest that 
the effect of the background energy distribution
is negligible in the case of a high  performance tracking device but can be significant if the angular resolution and/or sense recognition efficiency are poor. Indeed, as one can see
from the red dotted line in figure \ref{fig:BackSlopeProspect}, the sensitivity is degraded by almost one order of magnitude when the WIMP mass corresponds to the 
considered value of $E_{\rm back}$ and if the detector does not have sense recognition.

\section{Prospect}
\label{sec:prospect}

In this last section, we are interested in evaluating the discovery potential of two very different directional experiments referred to as  
  detector A and B, see 
Tab.~\ref{tab:detectors}. Both of them are considered as 10 kg of CF$_4$ Time Projection Chambers operated during three years with three dimensional track 
reconstruction. High performance characteritics are considered for detector A, while detector B stands for the pessimistic case.
Indeed, the detector A will be considered with an energy threshold of 5 keV, with an angular resolution of
20$^{\circ}$, a 100\% sense recognition efficiency and no background contamination; while detector B is characterized by an energy threshold of 20 keV, an angular resolution
of 50$^{\circ}$, no sense recognition capability and a high background contamination of 10 events/kg/year uniformly distributed in energy.
 This way, detector B can be considered as a pessimistic scenario
and detector A as an optimistic one that could also be interpreted as the ultimate directional detector.\\

 \setlength{\tabcolsep}{0.1cm}
\renewcommand{\arraystretch}{1.4}
\begin{table}[h]
\begin{center}
\hspace*{-0.5cm}
\begin{tabular}{|c||c|c|c|c|}
\hline
  & $E_{\rm th}$ [keV]  &  $R_{b}$ [evts/kg/year] & $\sigma_{\gamma}$ [$^{\circ}$] & $\epsilon_{\rm HT}$ [\%] \\ \hline \hline  
 Detector A & 5 & 0 & 20 & 100 \\ \hline 
 Detector B & 20 & 10 & 50 & 0 \\ \hline 
\end{tabular}
\caption{Characteristics of the two detector A and B considered.}
\label{tab:detectors}
\end{center}
\end{table}
\renewcommand{\arraystretch}{1.1}

\begin{figure}[t]
\begin{center}
\includegraphics[scale=0.4,angle=0]{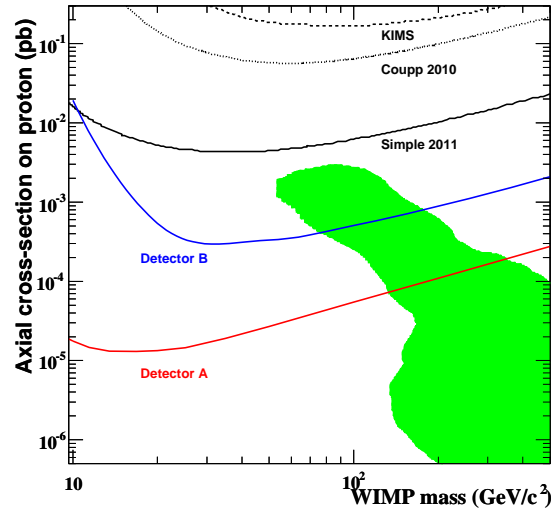}
\caption{Lower bound of the 3$\sigma$ discovery region at 90\% C.L. in the ($m_\chi,\log_{10}(\sigma_p)$) plane for the two detectors A (red solid line) 
and B (blue solid line), considering a 30 kg.year exposure.
The theoretical region, obtained within the framework of the constrained minimal 
 supersymmetric model, is taken from \cite{superbayes} and shown as the green contour. Constraints from collider data and relic abundance are accounted for.
Exclusion limits from direct searches are also shown: SIMPLE \cite{simple} (black solid line), COUPP \cite{coupp} (dotted line) 
and KIMS \cite{kims} (dashed line).} 
\label{fig:ProspectsFinal}
\end{center}
\end{figure}

On figure \ref{fig:ProspectsFinal} we present  the  
lower bound of the 3$\sigma$ discovery region at 90\% C.L. in the ($m_\chi,\log_{10}(\sigma_p)$) plane corresponding to the detectors A (red solid line) and B (blue solid line). For comparison
with other direct searches, we have reported current limits from 
SIMPLE \cite{simple} (black solid line), COUPP \cite{coupp} (black dashed line) and KIMS \cite{kims} (black dotted line). All limits are given in the pure-proton approximation 
\cite{tovey}. To evaluate the discovery potential of Dark Matter corresponding to the detectors A and B we have reported the theoretical region 
obtained within the framework of the constrained minimal supersymmetric model, taken from \cite{superbayes}, as the green contour.\\

The first result that can be inferred from figure \ref{fig:ProspectsFinal} is that the sensitivity of detector A is about one order 
of magnitude below the one from detector B for WIMP masses above 50 GeV/c$^2$ and about three order of magnitude for $m_{\chi} = 
10$ GeV/c$^2$. This highlights the need for detector optimization and justify a substantial experimental effort on directional 
detection to reach a high performance detector.
It also means that, contrarily to detector A, detector B will not be sensitive
to light WIMP below $\sim$ 20 GeV/c$^2$. The second result is that both detector A and B are competitive as 
they are 1 and 2 orders of magnitude below current exclusion limits. Moreover, both of them should be able to reach a 
large part of the predicted supersymmetric model which is motivated both by particle physics and cosmological
constraints. Hence, even a low performance directional detector, with a 30 kg.year exposure, could allow to identify WIMP events as such by recovering, 
with a high significance, the main incoming direction of the events.

\section{Conclusion}

In this paper, we have shown that the use of a profile likelihood ratio test statistic is very well suited when trying to estimate the sensitivity of a given
directional experiment. Indeed, it allows us to propagate, using a frequentist approach, some astrophysical uncertainties that are very important in the field of
direct searches of Dark Matter. This way, we have studied one by one the impact on the sensitivity of most of the experimental issues related to directional detection in the
goal of directional detection optimization. We
have then been able to elaborate a weighted wish list where we found that the energy threshold and the background contamination are the most dominant effect. About the effect
of angular resolution and sense recognition efficiency, we found that they could affect strongly (about a factor of 4-5) the sensitivity of a directional experiment
especially at high WIMP mass $m_{\chi} > 100$ GeV/c$^2$. On the other hand, we found 
that the energy resolution of the detector affects the sensitivity only in a negligible way, even in extreme and unrealistic cases. However, 
detector commissioning is compulsory to derive relevant exclusion limits or discovery regions.\\
Finally, we found that directional detection with a 10 kg CF$_4$ Time Projection Chamber should be able to reach an important part
 of   supersymmetric models and be
competitive with current experimental limits, even in the case of a directional experiment with very low performance. Hence, we believe that directional detection
of Dark Matter is a very promising direct search for Dark Matter that should be able to clearly authenticate a genuine positive 
detection with a high significance.

\section*{Acknowledgements}
J. B. would like to thank Benoit Clement and Glen Cowan for fruitful discussions and valuable advice concerning the profile likelihood ratio test statistic.

\end{document}